\documentclass[useAMS,usenatbib]{mn2e}
\usepackage{epsfig}
\usepackage{amsmath}
\usepackage{rotating}
\usepackage{multirow}
\usepackage{natbib}
\usepackage{longtable}

\newcommand{\be}{\begin{equation}}
\newcommand{\beq}{\begin{equation}}
\newcommand{\ba}{\begin{eqnarray}}
\newcommand{\ee}{\end{equation}}
\newcommand{\eeq}{\end{equation}}
\newcommand{\ea}{\end{eqnarray}}

\def\lsim{~\rlap{$<$}{\lower 1.0ex\hbox{$\sim$}}}

\def\gsim{~\rlap{$>$}{\lower 1.0ex\hbox{$\sim$}}}

\begin{document} 

\title[Emission Line Profiles with Resonant Scattering]{Disentangling Resonant
  Scattering and Gas Motions in Galaxy Cluster Emission Line Profiles}

\author[C. Shang and S. P. Oh]{Cien Shang$^{1}$ and S. Peng Oh$^{2}$\\
$^{1}$Kavli Insitute for Theoretical Physics, University of California, Santa Barbara, California, CA 93106; cshang@kitp.ucsb.edu\\
$^{2}$Department of Physics, University of California, Santa Barbara, California, CA 93106; peng@physics.ucsb.edu}

\date{\today}

\maketitle

\label{firstpage}

\begin{abstract}
Future high spectral resolution telescopes will enable us to place
tight constraints on turbulence in the intra-cluster medium
through the line widths of strong emission lines. At the same time, these bright lines are the most prone to be optically thick. 
This requires us to separate the effects of resonant scattering from
turbulence, both of which could broaden the lines. How this can be achieved has yet not been quantitatively addressed. In this paper, we
propose a flexible new parametrization for the line profile, which allows these effects to be distinguished. The model has only 3 free parameters, which we calibrate with Monte-Carlo
radiative transfer simulations. We provide fitting functions and tables that allow the results of these calculations to be easily incorporated into a fast spectral fitting package. In a mock spectral fit, we
explicitly show that this parameterization allows us to correctly estimate the
turbulent amplitude and metallicity of a cluster such as Perseus, which would otherwise give significantly  
biased results. We also show how the physical origin of the line shape can be understood analytically. 
\end{abstract} 

\section{Introduction}
\label{sec:introduction}

Turbulence plays a key role in many
aspects of the physics of the intra-cluster medium (ICM). It provides
non-thermal pressure support, thus biasing the masses
measured in X-ray \citep{lau09} and via the 
Sunyaev-Zeldovich
effect\citep{shaw10,battaglia11,battaglia11a,parrish12}, which assume thermal pressure alone.  This uncertainty in cluster mass calibration constrains the use of clusters for precision cosmology. Turbulence could also transport metals \citep{rebusco05,
simionescu08}, amplify magnetic fields \citep{subramanian06,
  ryu08, cho09, ruszkowski11a}, and by accelerating cosmic rays, is likely to be closely related to radio
phenomena in clusters \citep{brunetti01, brunetti07}. Moreover,
turbulence can play a crucial role in the thermal state of the
cluster. It can halt catastrophic cooling by dissipation of 
turbulence \citep{churazov04, kunz11}, turbulent heat diffusion, 
\citep{cho03, kim03, dennis05}, or more subtly, by randomizing 
the magnetic field, and restoring 
thermal conduction to $\sim 1/3$ of the Spitzer value
\citep{ruszkowski10, ruszkowski11, parrish10}. Amongst others, gas motions can be seeded by structure formation (including motions of sub halos and mergers) and AGN activity. 

Despite its importance, turbulence in the ICM has so far only been weakly
constrained, generally by indirect means. These have came
from the analysis of pressure maps \citep{schuecker04}, surface
brightness fluctuations\citep{Churazov2012, Sanders2012}, the lack of
detection of resonant-line scattering \citep{churazov04,werner10},
Faraday rotation maps \citep{vogt05,enslin06}, and deviations from
hydrostatic equilibrium with thermal pressure alone
\citep{Churazov2008,churazov10,Zhang2008}. Constraints can also be 
provided by a more direct means -- measuring the Doppler broadening
of strong emission lines. Unfortunately, due to the poor energy
resolution of current X-ray telescopes, to date this type of analysis has only yielded
upper limits \citep{Sanders2010, sanders11, dePlaa2012}. The challenge stems also from the fact that the expected gas motions are highly subsonic, with Mach numbers ${\cal M} \sim 0.1-0.3$ in the cluster core.  

However, this situation is expected to change with the launch of
Astro-H\footnote{http://astro-h.isas.jaxa.jp/} in 2014 and 
ATHENA\footnote{http://sci.esa.int/ixo} in the farther future. With their high 
spectral resolution (e.g., ${\rm FWHM} \sim 4-5$ eV for Astro-H), these telescopes can accurately
measure the widths of emission lines, thereby placing tight
constraints on the amplitude of turbulence. They may even allow us to
separate different modes of gas motions by mixture modeling of the detailed line
profile \citep{Shang2012}. For instance, bulk motions and gas sloshing can be identified through components with different means, while the volume filling fraction of turbulence can be identified through components of different widths. To some extent, the high spectral resolution of Astro-H can mitigate against its poor spatial resolution in learning about the detailed velocity field. 

To correctly recover the amplitude of turbulence from emission lines, one
important factor has to be taken into account: resonant
scattering. The ICM is optically to 
thin continuum free-free emission. However, the optical depths for a few
strong metal emission lines could still exceed unity \citep[and references
therein]{Churazov2010a}. Photons in the lines therefore experience
multiple scatterings before they reach the edge of the cluster. Such
RS has many important consequences. 
For instance, the line ratios of optically thick and thin lines are
altered. The line width and shape are changed as well, since photons
in the optically thick line center are scattered to the optically thin line wings, where they can escape. Morever, these effects happen systematicaly as a function of position, since photons escape more easily from the cluster outskirts than from the center. Under certain
conditions, the variance of the line could increase by 40 \%. Failure to
account for resonance scattering introduces biases in the measurement of the
metallicity \citep{Sanders2006} and turbulence (as we shall show). 

RS in the context of galaxy clusters has been studied both
analytically and with radiative-transfer Monte-Carlo simulations. It was first considered by \citet{Gilfanov1987},
who discussed various features of the RS effect and
their astrophysical implications. \citet{Mathews2001} considered if the RS could sufficiently explain the abundance hole of metals
seen in M87. \citet{Sazonov2002} demonstrated that Rayleigh scattering
could polarize the line emission, and detection of the polarized
X-rays could yield valuable information on ICM physical
conditions. This work was later followed up by \citet{Zhuravleva2010},
who proposed to use the polarization of X-ray lines as a way of
measuring the tangential component of the ICM
velocity. \citet{Molnar2006} discussed how the resonant scattering
lines could be used to measure angular diameter distances to clusters
and consequently constrain cosmology. \citet{Sanders2006} considered the effect of RS on the inferred metallicity of the cluster, via an analytic calculation which can used to fit observations. This work is most similar in spirit to this paper, although they adopted the single scattering approximation, ignored gas motions, and did not include the line broadening due to RS. Most recently,
\citet{Zhuravleva2011} investigated how the RS depends on the spatial
scale and anisotropy of gas motions and how such dependence could be used
to probe the characteristics of gas motions. 

Despite the increasing sophistication of these studies, perhaps the most important leading order effect--the joint effects of turbulence and RS in altering the line profile---has not been quantitatively addressed. Studies have generally either considered the effects of RS in a static ICM, or the effect of gas motions in an optically thin ICM. In reality, the gas is likely to always have subsonic motions, and the brightest lines with the highest oscillator strengths (e.g., the He-like line of Fe XXV) we will use to measure these motions are by the same token the most likely to be optically thick, and the effects of RS cannot be ignored. It is important to consider these effects in tandem for two reasons. Firstly, as previously mentioned, by altering line widths, RS could bias measurements of turbulence. Our first purpose in this paper is {\it to learn if one can strip away the confounding effects of RS to learn about the underlying velocity field.} Secondly, the two effects interact with one another: turbulence reduces the line optical depth and hence the effects of RS, while RS alters the effective velocity shift experienced by photons. Our second purpose is therefore a practical one: {\it we aim to provide observers a parametric model of the 
  line profile which takes both RS and turbulence into account, in order to quickly and accurately estimate turbulent broadening.} 
The alternative would be to run radiative transfer simulations for each plausible realization of a given cluster's velocity field, a very costly and slow way of fitting the data. 

The remainder of this paper is organized as follows. \S~
\ref{sec:methodology} describes the assumed cluster model, the
line profile model and procedures of Monte-Carlo simulation and
profile fitting. \S~\ref{sec:result} discusses how the line profile
depends on various physical parameters. We then calibrate the
dependence with a suite of simulations. The results are given in both
tabulated and analytical fitting function form. We also test the model and the
calibration by comparison with a new set of simulations, 
performing a mock spectral fitting. \S~\ref{sec:analytic} describes how the physical origin of the characteristic line shape after resonant scattering can be understood analytically. 
\S~\ref{sec:summary} summarizes our 
results and discusses possible extensions to the work.

\section{Methodology}
\label{sec:methodology}

In this section, we describe our method of a flexible parametric model
to describe the line profile in the presence of gas motions and
resonant scattering. We assume that the cluster temperature, density and metallicity\footnote{We justify this assumption later.} profiles are known ahead of time from X-ray observations, and only the velocity field is unknown. Given realizations of the velocity field motivated by numerical simulations, we fit a line profile which separates the effects of turbulent broadening from that of RS.  

\subsection{Cluster Model}
\label{subsec:clustermodel}

{\bf Electron density profile} The isothermal $\beta$ model has been widely used to describe the free-free surface
brightness and density profiles of clusters
\citep{Cavaliere1976}, 
\ba
n(r)=n_0 \left(1+\frac{r^2}{r_c^2}\right)^{-3\beta/2},\\
S(r)=S_0 \left(1+\frac{r^2}{r_c^2}\right)^{1/2-3\beta},
\label{eqn:bmodel}
\ea
Here $n_0$ and $S_0$ are the central density and surface brightness,
respectively, and $r_c$ is the core radius, while $\beta$ controls the asymptotic density profile at large radii. 
The model is very simple with only three free parameters.
Overall, it describes observed profiles reasonably well, although the
density is often over(under)-predicted at the large(small)
radii. For this reason, several extensions have been proposed,
which fit observed profiles better at the cost of more free parameters
\citep[e.g.,][]{Chen2007, Vikhlinin2006}. 


In this first attempt to systematically model the line profile
with resonant scattering, we restrict ourselves to the simple
$\beta$ model. Since the line optical depth is only linearly proportional to
density, we do not expect the results in this paper to be strongly affected by deviations from the $\beta$ model. 

{\bf Ion density profile} Resonant scattering ($\propto n_{\rm Z_{i}}$) and line emission ($\propto n_{\rm Z_{i}} n_{\rm e}$) is sensitive to the density of metal ions $n_{\rm Z_{i}}$. 
This is sensitive to two factors: metallicity ($n_{\rm Z} \propto Z$), and temperature (which sets the ionic abundance in coronal equilibrium). In this paper, we shall also assume that $n_{\rm Z_{i}}$ can be fit by a $\beta$ profile (with different parameters from the electron profile). While not true in detail, it is reasonably accurate for the central core where most of the emission and resonant scattering takes place; the beta profile can be viewed simply as a mathematical fitting function for $n_{\rm Z_{i}}$ with a central core and asymptotic outer slope, without reference to the electron $\beta$ profile. In practice, $n_{\rm Z_{i}} \propto n_{e}$ is a good approximation for NCC clusters, as we shall verify\footnote{This arises simply because the gas is approximately constant metallicity and isothermal in the core of NCC clusters, and the first order variations (declining metallicity and decreasing temperature with radius) tend to cancel out.}. Since this is mathematically simpler, we treat this case first. We consider the NCC case where $n_{\rm Z_{i}},n_{\rm e}$ have different $\beta$ model parameters in \S\ref{subsec:application}. 

{\bf Temperature} The temperature profile has two effects: affecting ionic abundance in coronal equilibrium (discussed above), and making thermal broadening a function of radius. The former effect is the dominant one. For the latter, we initially assume spatially constant thermal broadening (i.e., isothermal temperature, which is approximately true in the core), and then show that taking the true temperature profile into account has little effect. 

{\bf Velocity Field} We generate realizations of turbulence over a 3D grid of
$256^3$, with a box size of 2.2 Mpc. The velocity fields are generated
from a power spectrum of the form: 
\ba
E(k) =
\left\{\begin{array}{ccc}
N_0& k\le k_\mathrm{c}\\
N_0\left(\frac{k}{k_\mathrm{c}}\right)^{n_\mathrm{s}}& k>k_\mathrm{c}
\end{array}\right.,
\label{eqn:velocity_PS} 
\ea
This form is motivated by both simulations and indirect
observations 
of pressure fluctuations which find power-law like spectra and a cutoff in the
large scale \citep[e.g.,][]{Schuecker2004, Vazza2011, Vazza2012}. 
For Kolmogorov turbulence, the slope $n_\mathrm{s}=-5/3$.

The use of a power spectrum alone to describe turbulence is likely to be incomplete and over-simplified,
since it does not take possible spatial dependence of the gas motions into account, particularly radial dependence as well as anisotropy (which is important since we can only probe gas motions along the line of sight). Strictly speaking, the velocity field in a cluster is likely to be non-Gaussian. 
A more realistic approach would be to take the velocity fields directly from
numerical simulations. However, this simple prescription is in relatively good agreement with numerical simulations, and moreover allows us to rapidly parameter space. For relatively simple statistics such as line height and width, a power spectrum parametrization is adequate. This is less likely to be true for higher order statistic which depend on details of the line shape \citep{shang12}. 

Another possible concern lies in the assumption that turbulence does not exhibit systematic spatial trends in the cluster. This seems in contradiction with
recent simulation results \citep[e.g.,][]{lau09, battaglia11}, which
show a power law increase toward large radii. However, the power law
increase usually takes place around $0.5-1r_{500}$ (0.5-1 Mpc for
nearby bright clusters), which is
considerably larger than the core radius ($\lsim$ 200 kpc). We
therefore do not expect it to significantly alter our conclusions.


The normalization factor $N_0$ is adjusted to obtain the desired RMS
velocity. From numerical simulations, we generally expect turbulent
gas motions to be highly subsonic, with Mach numbers ${\cal M} \sim
10-30\%$, although the Mach number rises as a function of radius and
can be transonic at the virial radius \citep{shaw10}.  
We also place an upper bound on sub-grid turbulence by integrating the power spectrum from $k_{\rm max}$ (the
wavenumber corresponding to the smallest cell; in our simulations the Nyquist length scale is 8.6 kpc) 
to infinity. The
velocity dispersion of this sub-grid turbulence, $\sigma_{\rm sub}$, will be used in the following steps.

\subsection{Line Profile Model}
\label{sec:model}

Resonant scattering (RS) modifies the line profile by redistributing photons from the optically thick line center to the optically thin line wings, where they can escape to the observer. Here, we describe our fitting function to the line shape.

In the absence of resonant scattering, the emission line profile can be well
approximated by a Gaussian, so long as the injection scale of
turbulence 
is smaller than the emitting region where most photons arise (thus, the field of view encompasses many independent `patches' and the central limit theorem holds). 
RS modifies the line
shape due to its frequency-dependent cross section $\Sigma$ (the symbol $\Sigma$ is used to distinguish it from the line dispersion $\sigma$),
\ba
\Sigma(E, \sigma) \propto \frac{1}{\sqrt{2 \pi} \sigma} e^{-\frac{(E-E_0)^2}{2\sigma^2}},
\ea
where $E_0$ is the central energy of the line and $\sigma$ is the
dispersion of the line, including contributions 
from both thermal and turbulent motions: 
\ba
\sigma^2=\sigma^2_{ther}+\sigma^2_{turb}=\frac{E_0^2}{c^2} (\frac{k T}{A
  m_p} + \sigma_{\rm turb}^2).
\ea
Here, $\sigma_{\rm turb}$ is the 1D RMS velocity dispersion,
$c$ is the speed of light, $m_p$ is the proton mass, and $A$ is the
atomic number. In the discussions that follow, we specialize to the iron ion ($A=56$), in particular the He-like 6.7 keV 
line, which for instance in the absence of turbulent motions has a line center optical depth of 2.7 in Perseus and 1.4 in Virgo \citep{Churazov2010}. 
However, our methods could be applied to any other resonance line. 
Photons in the optically thick line center are scattered out of the line of sight, while photons in the optically thin line wings escape freely toward the observer. During the scattering process, photons which are re-emitted at line center can continue to scatter, while photons which are re-emitted in the lines escape. Thus, the line center is suppressed, while the wings are slightly enhanced due to re-emission. This is clearly seen in 
Fig. \ref{fig:example}. 

We therefore model the line profile as:
\ba
P(E, \sigma)=A(E, \sigma) R(E, \sigma) G(E, \sigma),
\label{eqn:pdf0}
\ea
where $G(E, \sigma)$ is the original Gaussian line profile without RS, and $A(E,
\sigma)$ and $R(E, \sigma)$ are factors 
accounting for absorption and reemission, respectively.
We found the modification to the line profile due to re-emission 
could be adequately modeled with a simple power law with two free parameters: 
\ba
R(E, \sigma)=(1+g_1 \tau_0^{h})= (1+g_1 (\zeta/\sigma)^{h}),
\label{eqn:pdf1}
\ea
where $\tau_0$ is the opacity at line center from the cluster center to
infinity, while $\zeta \equiv \tau_0 \sigma$ (since $\tau_0 \propto \Sigma \propto \sigma^{-1}$, we set $\tau_0=\zeta/\sigma$). We separate $\sigma$ from $\tau_0$ in the equations, as it is the
quantity of interest in this paper. 
$\zeta$ can be readily computed in terms of fundamental constants and observed quantities: 
\ba
\zeta= \sigma \int_0^{\infty}\Sigma(E=E_0) n_i\text{d}l
=\int_0^{\infty}\sqrt{\pi} h r_e c f \delta_i (l) n_{\rm Z} (l)\text{d}l,
\label{eqn:tau}
\ea
where $h$ is the Planck constant, $r_e$ is the classical radius of the 
electron, $f$ is the oscillator strength of the transition, $\delta_i$
is the fraction of the element in the appropriate ionization state,
and $n_{\rm Z}$ is the number density of the element. 
We model the energy-dependent absorption using the usual
exponential function with an additional free parameter,
\ba
A(E, \sigma)=\text{exp}(-g_2 \tau(E))=\text{exp}\left(-g_2 \frac{\zeta}{\sigma}
    e^{-\frac{(E-E_0)^2}{2\sigma^2}}\right).
\label{eqn:pdf2}
\ea
In total, there are three free parameters: $g_{1},g_{2},h$. We find this model is sufficiently
flexible to fit the simulated profiles.

\subsection{Monte Carlo Radiative Transfer}
\label{subsec:mc}

For resonance line radiative transfer where multiple scatterings are possible, Monte Carlo simulations offer the greatest flexibility and ease of computations, and are standard in the field (e.g., \citet{bonilha79, ahn01, zheng02, hansen06}). The simulations performed for this work largely follow \citet{Zhuravleva2011a} (see
also \citet{Sazonov2002, Churazov2004, Zhuravleva2010}). For
completeness, we nevertheless describe briefly the main
procedures. A simulation is composed of the following steps:

{\bf 1) Initialization of photons.} 

We produce a mock photon with a unit weight $w$, propogating in a random
direction. Its location is randomly drawn according to the volume
emissivity distribution, while its energy is taken from a Gaussian
distribution with a variance of $\sigma^2_\mathrm{ther}+\sigma^2_\mathrm{turb}$. 
The mean of the Gaussian distribution takes into account the Doppler
shift due to cell motion, 
\ba
E_\mathrm{mean}=E_0 (1+{\bf v_{cell}{\cdot}m}/c),
\ea
where ${\bf v_{cell}}$ and ${\bf m}$ are the vectors of the cell velocity and
photon propogation direction, respectively. We record the initial
location and energy of the photon. This information is used to
extract line profiles without RS, which can be compared to
profiles with RS to determine model parameters.
 
{\bf 2) Photon propagation.}

Next, we integrate along the propagation direction ${\bf m}$ to find the optical
depth to the edge of the cluster and the corresponding escape 
probability $p_{esc} = e^{-\tau}$ (the cluster is cut off at 1
Mpc). The weight of the escaped photon is then equal
to $w p_{esc}$. Alternatively, the photon might be scattered. The
optical depth to the next scattering location, $\tau_{next}$, follows
an exponential distribution between 0 and $\tau$. We therefore randomly draw
$\tau_{next}= -\text{ln} (1 -\xi (1-p_{esc}))$, where $\xi$ is a
uniformly distributed random number between 0 and 1. The position of
the next scattering is then identified using $\tau_{next}$ and ${\bf m}$.

{\bf 3) Scattering.}

RS takes place and changes the direction of photon motion and its energy.
The new propogation direction of the scattered photon ${\bf
  m^{\prime}}$ is randomly drawn according to the proper scattering phase
matrix. There are two common types of scattering: isotropic and
Rayleigh (dipole) scattering. The scattered direction of the former is uniformly
distributed, while the probability of ${\bf m^{\prime}}$ for the latter
is proportional to $({\bf m\cdot m^{\prime}})^2$. Their relative weights depend on the total angular momentum change $j$ of the ground state and the difference $\Delta j$ between the ground and excited state. For the He-like 6.7 keV like of iron, scattering is purely dipole. Thus, in this paper we
mainly focus on Rayleigh scattering, while results for the isotropic
scattering will be posted on the internet\footnote{https://sites.google.com/site/cienshang/toolbox}.
To find the energy 
of the scattered photon $E^{\prime}$, we first need to compute the velocity of
the scattering ion. Along the original propogation direction ${\bf m}$, the
velocity of ion must satisfy $E_0=E(1-v_{ion,||}/c)$ in order for the
RS to happen. Perpendicular to ${\bf m}$, the ion velocity is drawn from 
a Gaussian distribution of mean ${\bf v_{cell}}$ and variance
$\sigma^2_\mathrm{ther}+\sigma^2_\mathrm{sub}$. The new energy is then computed using ${\bf
  m^{\prime}}$ and ${\bf v_{ion}}$, $E^{\prime}=E_0 (1+{\bf
  v_{cell}\cdot m^{\prime}})$. The scattered photon has a weight
$w^\prime=w(1-p_{esc})$. 

{\bf 4) Repeat} 

We repeat step 2-4 many times until the weight of
the photon is neglible ($<10^{-8}$). We run the simulation for
a large number of photons $(10^5-10^6)$, which allows us to build up good
statistics to compute the PDFs and fit the model parameters.

\subsection{Fitting Procedure}
\label{subsec:fitting}

We fit the simulated line profiles to obtain the best-fit parameters and their uncertainties. As these fits are performed many times, we
choose a simple and fast strategy: we minimize the $\chi^2$ between the model and binned data using the Levenberg-Marquardt algorithm implemented in the GSL scientific
library\footnote{http://www.gnu.org/software/gsl/}. 
In particular, we minimize:
\ba
\chi^2 = \sum_{i} \frac{(N_{sim,i}-N_{mod,i})^2}{N_{sim,i}},
\label{eqn:chisq} 
\ea
where $N_{sim,i}$ and $N_{mod,i}$ are simulated and model-predicted
photon counts in $i$-th bin, respectively. Since $\chi^2$ minimization is inaccurate for small
number statistics (where one should minimize the Cash C statistic \citep{Cash1979}, which correctly maximizes the Poisson likelihood), only bins with $N_{sim} >10$ are used in the fit. 

\begin{figure}
\begin{tabular}{c}
\rotatebox{-90}{\resizebox{60mm}{!}{\includegraphics{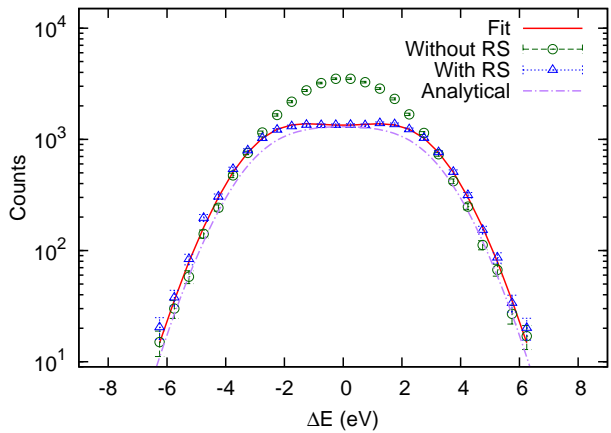}}}
\end{tabular}
\caption{Line profiles with and without RS from the Monte Carlo
  simulation, together with our best fit 
model. The photons are accumulated from ring with a width of 10 kpc,
centered at 55 kpc. Please see text for other details; the analytic profile is discussed in \S~\ref{sec:analytic}.}
\label{fig:example}
\end{figure}


\section{Results}
\label{sec:result}

The main task of this paper is to present a new model for the line
profile with RS, and to calibrate the dependence of the 
model on various physical quantities. We first consider the case when $n_{\rm Z_{i}} \propto n_{e}$, since this is mathematically simpler, and discuss the case where $n_{\rm Z_{i}}$ has a separate profile later.  

The relevant quantities are: the cluster temperature profile $T(r)$, $\beta$ model
parameters for the electron and ion profiles ($n_0$, $\beta$, $r_c$) -- note that the beta model parameters for the ions implicitly folds in information about the metallicity and temperature profiles), velocity field parameters
($\sigma_{turb}$, $n_{s}$, $k_{c}$), and geometrical parameters  
parameters (radius $r$, and field-of-view (FOV) of the
telescope). Dependence on some of the parameters has  
already been included in the model (equation \ref{eqn:pdf0} -
\ref{eqn:pdf2}) itself. For example, the dependence on temperature and
$\sigma_{turb}$ is approximately incorporated\footnote{This is exactly true if the velocity and temperature profiles are constant, and approximately true otherwise. See \S~\ref{subsec:powerspectrum} for more details.} through
$\sigma$:
both higher temperature and stronger turbulence raise 
$\sigma$. Likewise, the normalization of the ion profile, $n_{i,0}$ is set by $\zeta$ (see equation \S\ref{eqn:tau}). In the simulations, we fix $T$ to 4 keV
($\sigma_{ther}\sim 1.8$ eV
for iron ions)  
and set $\zeta=5$ eV (see discussion below). We then fit the line profile for various values of $\sigma_{turb}$ (and consequently $\sigma$). If our proposed model accurately fits the line profile, and factors $A(E,\sigma),R(E,\sigma)$ appropriately describe the effects of resonant scattering and can be accurately recovered, then the original line profile $G(E,\sigma)$ unmodulated by the effects of resonant scattering can be recovered. 

In order to reduce the dimensionality of the problem, it is useful to exploit self-similarity. There are 4 length scales in the problem: the photon mean free path at line center from the cluster center $\lambda$,  the projected radius $r$, the field of view FOV, and the core size $r_c$. The solution only depends on ratios of these length scales; for convenience, we choose to fix the core size $r_c$, and vary $\lambda, {\rm FOV}, r$. For definiteness, we set $r_{c}=100$ kpc, though obviously our solutions are independent of this choice, and would hold with appropriate rescaling. The photon mean free path $\lambda$ depends on the optical depth at line center $\tau_{0}$. Since $\tau_{0} = \zeta/\sigma$, we can choose to vary the overall normalization of the optical depth profile $\tau_{0}$ by fixing $\zeta$ and varying $\sigma$. Finally, given a model of the radially varying line profile collected in an infinitesimal area
$P(r)$, the observed profile is given by integrating the emission-weighted profile $P(r)$ over the FOV: 
\ba
P(FOV)=\frac{\int_{FOV} S(r) P(r) d^2{\bf r}}{\int_{FOV} S(r) d^2{\bf r}}.
\label{eqn:beam}
\ea
Note $P(r)$ is collected in a 2D ring on the sky, not in 3D shells. We check that self-similarity indeed holds numerically in \S\ref{subsec:test}.  

The dependence on the remaining four parameters ($n_s, k_c, \beta$ and
$r$) are however not as 
simple, and need to be determined by Monte Carlo simulations. Note that $\beta$ and $r$
are known to us while $n_{s}$ and $k_{c}$ are generally not
known. We therefore discuss these two classes of parameters separately. 

\subsection{Dependence on $n_{s}$  and $k_{c}$}
\label{subsec:powerspectrum}

\begin{figure}
\begin{tabular}{c}
\rotatebox{-90}{\resizebox{60mm}{!}{\includegraphics{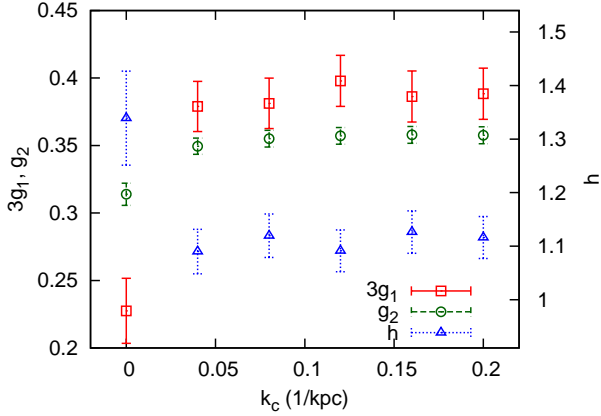}}}
\end{tabular}
\caption{Model parameters as functions of $k_c$. The photons are
  collected from a ring with a width of 10 kpc and inner edge
 at 100 kpc. $n_s$ is fixed to -5/3.}
\label{fig:kcut}
\end{figure}

\begin{figure}
\begin{tabular}{c}
\rotatebox{-90}{\resizebox{60mm}{!}{\includegraphics{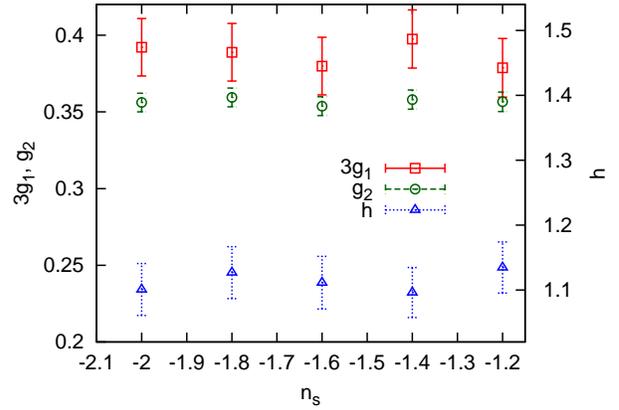}}}
\end{tabular}
\caption{Model parameters as functions of $n_s$. In this figure, $k_c$
is fixed to $0.1~{\rm
  kpc^{-1}}$.}
\label{fig:kslope}
\end{figure}

First, we run a series of simulations of $10^6$ photons with different $k_{c}$. In
these simulations, $n_{s}$ is fixed to its Kolmogorov value, $n_{\rm S}= -5/3$. We
collect photons from a set of concentric rings with width of $10$ 
kpc. This choice of width is narrow compared to $r_c$, and so can be
approximated as ``infinitesimal'' (the goodness of this approximation
will be checked in \S~\ref{subsec:test}), but still sufficiently wide 
to collect enough photons for good statistics. The inner edges of these
rings are at $r_{in}=0, 50, 100, 300$ kpc. The result for $r_{in}=
100$ kpc is shown in Fig. \ref{fig:kcut}, which shows a transition
around $k_{\rm c, crit} \sim 2\pi/r_c = 0.0628~{\rm kpc^{-1}}$. Before this transition,
the model parameters vary with $k_c$, while after it, they are
consistent with constant. The same conclusion could be drawn from
other cases. This result could be understood intuitively.
At small $k_c$, the outer stirring scale is larger than the size of the emitting
region ($\sim r_c$), and the LOS only ``sees'' part of an eddy. So the
motion is more similar to ordered bulk flow than chaotic. The
results in this regime depend on the detailed configuration of the
velocity field, and each case has to be studied individually (in particular, a mixture model approach to fitting the PDF \citep{shang12} may be applicable here). At
larger $k_c$, on the contrary, the eddy size is small relative to the
size of emitting region; since photons transverse many eddies, by the central limit theorem, the assumption of random motions which have a 1D gaussian profile is valid. 
In this regime, the line shape no longer depends on $k_c$ or
$\sigma_{ther}/\sigma_{turb}$.  Indeed, the model parameters are also independent of
$n_s$. This is seen in Fig. \ref{fig:kslope}, which shows
the parameters as functions of $n_s$, assuming $k_c=0.1~{\rm
  kpc^{-1}}$.Hereafter, we will focus only on this regime, and fix $k_c= 0.1~{\rm
  kpc^{-1}}$, and $n_s=-5/3$.
 
This relative independence to the details of the velocity power spectrum is easy to understand. Once $k_{\rm c} > k_{\rm c, crit}$, so that the velocity field sampled by outgoing photons is approximately Gaussian, then the velocity field largely affects the PDF by changing its line width (note that all these calculations are performed at fixed line width, by adjusting the power spectrum normalization $N_{0}$; equation \ref{eqn:velocity_PS}); the PDF is not sensitive to the detailed spatial structure of the velocity field. If we were to consider higher-order statistics which {\it are} sensitive to the spatial structure of the velocity field, such as the structure function \citep{zhuravleva12}, we would not observe this behavior. In this case, the independence of the details of the velocity field is a great advantage -- the effects of resonant scattering would otherwise be much more difficult to compute. 

\subsection{Dependence on $\beta$ and $r$}
\label{subsec:calibrate}

\begin{figure}
\begin{tabular}{c}
\rotatebox{-90}{\resizebox{60mm}{!}{\includegraphics{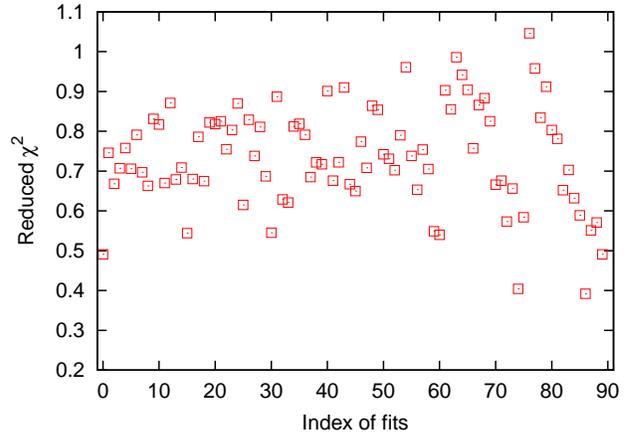}}}
\end{tabular}
\caption{Reduced $\chi^2$ for all the fits.}
\label{fig:chisq}
\end{figure}

To determine the dependence on $\beta$ and $r$, we run a suite of
simulations, varying $\beta$ from 0.5 to 1 with steps of 0.1, and $r$
(the radii of the rings from where the photons are collected)
from 5 kpc to 705 kpc with steps of 50 kpc. For a given set of $\beta$
and $r$, we set the r.m.s. of the turbulent velocity to 0, 100, 200, and
300 km/s, covering a range of $\tau_0$ from 2.7 to 0.7 (note that our
model automatically recovers the original Gaussian profile as $\tau_0
\rightarrow 0$, so the zero point does not need to be checked). Each set of $\beta$ and $r$ has 4 simulated 
profiles, which we fit simultaneously to constrain $p_1$, $p_2$, and
$h$. The results are tabulated in Table \ref{tbl:result} and \ref{tbl:result1} in the Appendix. An example is shown in Fig. \ref{fig:example} with $\beta=0.6$, $r=55$ kpc and
$\sigma_{turb}=0$. Fig. \ref{fig:chisq} shows reduced $\chi^2$ for all
the fits. The fits are
in general very good\footnote{The somewhat better than expected reduced $\chi^{2}$ likely comes from the simulation scheme of assigning weight to photons, which may artificially suppress the shot noise in the simulated profile.}, with the mean reduced $\chi^2=0.74$

Equation \ref{eqn:pdf0} - \ref{eqn:pdf2}, \ref{eqn:beam}, and Table 
\ref{tbl:result} are everything one needs to build a line profile model
for spectral fitting. For an arbitrary $\beta$ and $r/r_c$,
interpolation needs to be performed to determine the proper $p_1$,
$p_2$, and $h$. Alternatively, we provide below polynomial fitting
forms of $p_1(\beta, r)$, $p_2(\beta, r)$ and $h(\beta, r)$, found
by scientific data mining software {\sc Eureqa
  Formulize}\footnote{http://formulize.nutonian.com/} for $r/r_c 
\le 4$: 
\ba
p_1&=&0.0818 + 0.0642\beta^2\left(\frac{r}{r_c}\right) + 0.0469\beta\left(\frac{r}{r_c}\right)^2 \\\nonumber
&&+ 0.00626\left(\frac{r}{r_c}\right)^3 + 0.00393\beta^2\left(\frac{r}{r_c}\right)^4 \\\nonumber
&&- 0.0317\beta\left(\frac{r}{r_c}\right)^3;\\
p_2&=&0.671 + 0.0389 \left(\frac{r}{r_c}\right)^3 + 0.0418\beta^2 \left(\frac{r}{r_c}\right)^2 \\\nonumber
&&- 0.145 \beta - 0.0789\left(\frac{r}{r_c}\right)^2 - 0.318\beta\left(\frac{r}{r_c}\right) \\\nonumber
&&- 0.00495\left(\frac{r}{r_c}\right)^4;\\
h&=&1.69 + 0.523\left(\frac{r}{r_c}\right)^3 + 0.0130\left(\frac{r}{r_c}\right)^5 \\\nonumber
&&- 0.417\beta - 0.717\left(\frac{r}{r_c}\right)^2 - 0.140\left(\frac{r}{r_c}\right)^4;
\label{eqn:form}
\ea

\subsection{Parameter Space Test}
\label{subsec:test}

In obtaining the results in the above, we have made several 
simplifications. For example, the photons are collected from rings
of finite width; particular values of
$\zeta$, $T$, $k_c$ and $n_s$ were adopted. We argued that our result are nonetheless general, because $\tau_{0}$ is the only important factor governing the modification of the line profile, and varying $\sigma$ at fixed $\zeta$ is equivalent to varying $\tau_{0}$; for a fixed ion profile, $T$ only changes $\sigma$, which we directly constrain; and we showed in \S\ref{subsec:powerspectrum} that for $k_{\rm c} > k_{\rm c, crit}$, our results are independent of the values of $k_c$ and $n_s$. Nonetheless, these assumptions are worth checking. To confirm their validity and further test our model, we compare our model
against an ensemble of 735 simulated profiles. Parameters are randomly
drawn from reasonable ranges: $k_c\in[2\pi/r_c, 20/r_c]$,
$n_s\in[-2, -1.2]$, $\zeta\in[2, 6]$, $T\in[2, 8]$ keV, $\beta\in[0.5, 1.0]$
and $v_{rms}\in[0, 500]$ km/s. Each simulation produces $2\times 10^5$
photons, which are collected over a set of rings at $r<7r_c$ (the
largest radius within which our model has been calibrated). The rings
have random widths between $0.25r_c$ and $4r_c$, a range roughly
corresponding to the half-power-diameter (HPD) of Astro-H relative to the
$r_c$ of nearby clusters (see Table 2 of \citet{Shang2012} for the HPD and
\citet{Chen2007} for the $r_c$). 

\begin{figure}
\begin{tabular}{c}
\rotatebox{-90}{\resizebox{60mm}{!}{\includegraphics{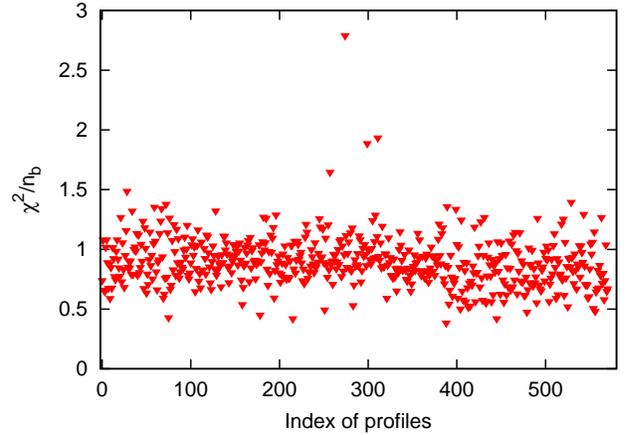}}}
\end{tabular}
\caption{$\chi^2$ normalized by the number of bins for the
  profiles in our test run.}
\label{fig:chisq2}
\end{figure}

\begin{figure}
\begin{tabular}{c}
\rotatebox{-90}{\resizebox{60mm}{!}{\includegraphics{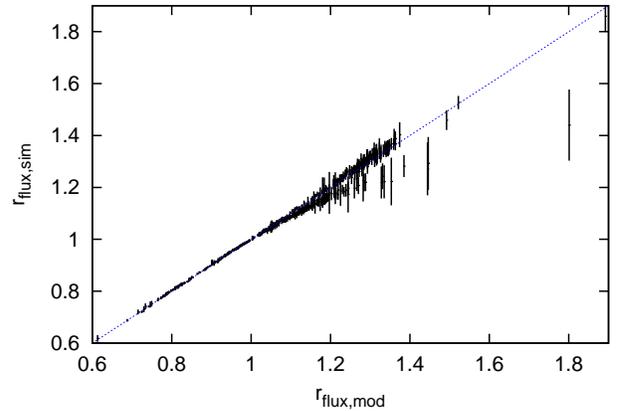}}}
\end{tabular}
\caption{Flux ratios from the simulation (y axis) and our model (x axis) between line
  profiles with and without RS. The straight dotted line is $r_{\rm flux,sim}=r_{flux,mod}$}.  
\label{fig:fluxratio}
\end{figure}

\begin{figure}
\begin{tabular}{c}
\rotatebox{-90}{\resizebox{60mm}{!}{\includegraphics{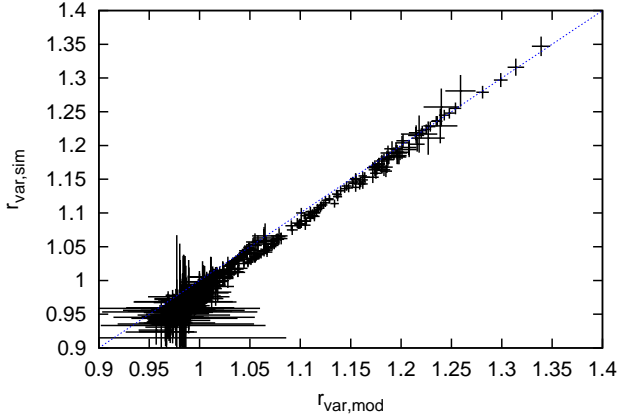}}}
\end{tabular}
\caption{Variance ratios between line profiles with and without RS. As
in Fig. \ref{fig:fluxratio}, x- and y- axes are the results from our
model and the simulation, respectively, while the dotted line is
$r_{\rm flux,sim}=r_{\rm flux,mod}$. }
\label{fig:varratio}
\end{figure}

Fig. \ref{fig:chisq2} shows the ratio of $\chi^2$ to the number of
bins $n_b$ for all the profiles in the test run. This ratio, similar
to the reduced $\chi^2$, is an indication of the goodness of fit. All
of our profiles have $\chi2/n_b\sim 1$, except for a few outliers which we
shall discuss later. The 
constraints on turbulence come primarily from the the height and
variance of the line. It is therefore crucial that our model
accurately reproduce these two quantities.
In Fig. \ref{fig:fluxratio}, we show the flux ratio between lines with
and without RS. The x- and y- axes are the values from our model and
the simulation, respectively, while the dotted line is where the model
exactly agrees with the simulation. We see that resonant scattering can both decrease the flux, $r<1$, by scattering photons from the cluster core out of the field of view, and increase the flux, $r>1$, by scattering photons from the core into the field of view.    
Overall, our model shows very good
consistency with the simulation, except for a few outliers. Similar
consistency is seen in Fig. \ref{fig:varratio} where we plot the
variance ratio between lines with and without RS. 

As for the a few outliers in Fig. \ref{fig:chisq2} and
\ref{fig:fluxratio}, careful examination reveals that they are all
related to emission at large radii ($r\gsim 6 r_c$). This likely
indicates that our calibration at large $r$ is inaccurate. Two reasons
might be responsible for this. First, due to the declining surface
brightness, the photon counts at large radii in our simulation are not
large enough to give us good statistics. Second, the large photon
collecting ring picks
up residual large scale motions in the velocity field. Note that the
velocity power spectrum did not go to 0 as $k \rightarrow 0$, but
stays as a constant. The parameter values in Table \ref{tbl:result} also show large scatter at large $r$. Our results for $r\gsim 6 r_c$ therefore cannot be trusted; however, this region contributes negligibly to an emission-weighted observation, and this shortcoming is of little importance.   

\subsection{Application to Clusters}
\label{subsec:application}
\begin{figure}
\begin{tabular}{c}
\rotatebox{-90}{\resizebox{46mm}{!}{\includegraphics{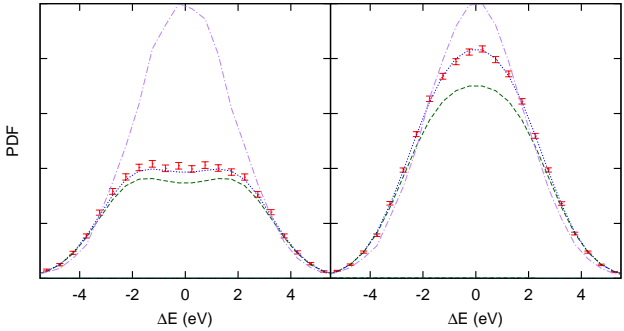}}}
\end{tabular}
\caption{Comparison between our model and simulations, when the core radii for electrons and ions are the same, but the outer profiles differ. The ion density is assumed to follow a $\beta$ model with
  $\beta=0.8$, while the electron distribution has $\beta=0.6$. The  dotted blue curves are model outputs when the ion profile is used as our model
  input while dashed green curves use the electron profile. The
  dot-dot-dashed curves are simulated PDFs without RS, while the data
  points are those with RS. The left and right panels are for the
  inner $0.5r_c$ and an annulus from $r_c$ to $2r_c$, respectively. Using ion profiles provides accurate results which agree with the full simulations.}
\label{fig:compare}
\end{figure}

\begin{figure}
\begin{tabular}{c}
\rotatebox{-90}{\resizebox{60mm}{!}{\includegraphics{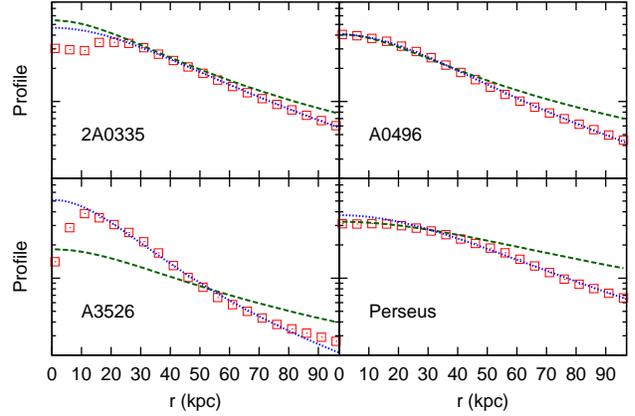}}}
\end{tabular}
\caption{Electron (dashed green curves) and Fe XXV ion densities (red squares) as a function of radius for four nearby bright clusters. The dotted blue curves are the best-fit beta model to the ion profile outside the cores. }
\label{fig:ccc_prof}
\end{figure}

\begin{figure}
\begin{tabular}{c}
\rotatebox{-90}{\resizebox{60mm}{!}{\includegraphics{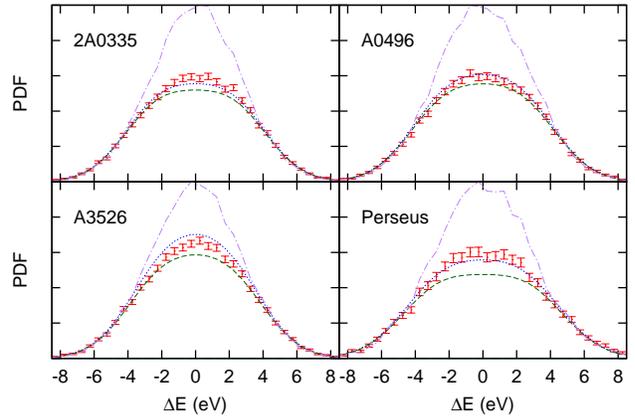}}}
\end{tabular}
\caption{Comparison of line profiles between our model and simulations for four nearby CCCs. The line and point types are same as in Fig. \ref{fig:compare}. The PDFs are from the central 30 kpc.}
\label{fig:ccc_pdf}
\end{figure}

\begin{figure}
\begin{tabular}{c}
\rotatebox{-90}{\resizebox{60mm}{!}{\includegraphics{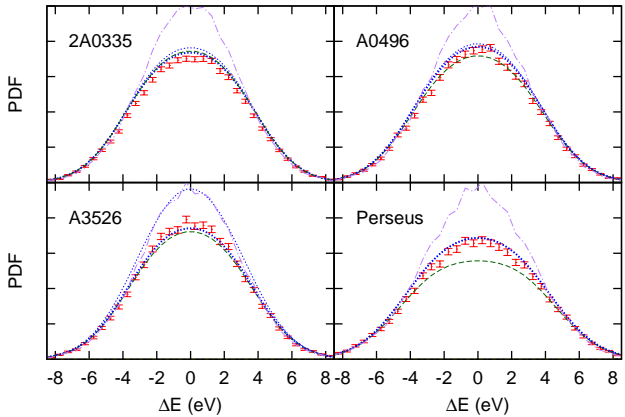}}}
\end{tabular}
\caption{Same as Fig. \ref{fig:ccc_pdf} except that the PDFs here are extracted from an annulus between $30-60$ kpc. {\bf The thick dotted blue curves are when the ``point source" correction is applied (see text for details).}}
\label{fig:ccc_pdf1}
\end{figure}

In the above, we performed calibrations for the case when the ions and electrons have identical $\beta$ profiles which differ only in normalization (which would require isothermal clusters of constant metallicity). However, distributions in real clusters deviate from such a idealized model, and will affect where the emissivity profile ($\propto n_{e} n_{i}$), which dictates where photons originate, and the optical depth profile ($\propto n_{i}$), which dictates where they scatter. We now discuss how these deviations affect our results, and how they can be corrected for.  

Most non cool-core clusters (NCCCs) can be reasonably be described by our model. This may seem somewhat surprising given that both of their temperature and metallicity profiles tend to decline toward large radii \citep[e.g.,][]{DeGrandi2001,Baldi2007,
  Leccardi2010}. However, the abundance of Fe XXV, the ion responsible for the
He-like iron line, declines with temperature at $T > \sim 3$ keV
(which is usually true for bright clusters). So the decline of temperature and
decline of metallicity have opposite effects which roughly cancel,
making the ion abundance roughly constant in the core. 

To conservatively estimate the uncertainty introduced by differences in the electron and ion profiles outside the core, we perform the following illustrative calculation. We assume
the ion density and electron density are described by two $\beta$
models with the same core radius but different $\beta$. According to
\citet{Baldi2007}, both the temperature and metallicity decline with a
power law slope of $\sim 0.3$ outside the core. Since $n \propto r^{-3\beta}$ for $r \gg r_{\rm c}$, we set $3(\beta_{\rm i} - \beta _{\rm e} = 0.6$ to account for this, or $\beta_{\rm i} - \beta_{\rm e} =0.2$. We then ran Monte Carlo radiative transfer calculations with these electron and ion profiles, and compared them to results from our fitting formulae. The results
are shown in Fig. \ref{fig:compare}. The left and
right panels show the PDFs from the inner $0.5 r_c$ and an annulus from $r_c$ to $2r_c$, respectively. The simulated PDFs (the data points) are compared
to our model with two sets of input parameters. The dashed green
(dotted blue) curves are when the $\beta$ model parameters for
the electron (ion) distribution are used. The fit is good when the beta model parameters for the ion distribution, rather than the electron distribution,
is used. This is reasonable, since the region outside the core is more important for scattering rather than producing photons, so here the optical depth profile is more important than the emissivity profile.  

However, the brightest nearby clusters which Astro-H will first turn to are cool-core clusters (CCCs). In CCCs, a $\beta$ profile can be a relatively poor fit to the ion distribution in the center. In particular, observed temperature and metallicity profiles often imply ion profiles which are non-monotonic and can decline, rather than remain constant, toward the center. This deviation will affect the number of scatterings taking place in the core and the number of photons emitted by the core but scattered at outskirts, affecting the profiles seen in the core and outskirts respectively. To estimate its importance, we perform calculations for four nearby bright CCCs -- 2A0335, A0496, A3526, and Perseus -- with measured temperature and  density distributions. The electron density distributions are taken from the $\beta$ model fit in \citet{Chen2007}, while the temperature and metallically distributions are from \citet{Snowden2008}. Three of the four clusters, except A0496, show dips in their metallicity distribution in the innermost bins, 
which might have been caused by RS. We thus ignore these bins, and instead assume the metallicity distribution is flat within the second innermost bin. We then compute their Fe XXV ion density profiles, based on these distributions and appropriate rate coefficients for coronal equilibrium \citep{Bryans2009}. They are shown in Fig. \ref{fig:ccc_prof} with squares. Compared to the electron density distributions (the dashed curves), the ion distribution is steeper outside the cores. Notably, the ion distribution in A3526 has a steep drop in the center, due to the temperature drop. Overall, the complex ion distributions cannot be fully described by a simple and general form. 

Nonetheless, to test the applicability of our previous calculations, we perform $\beta$ model fits to the profiles at $r>15$ kpc, shown with dotted blue curves. Monte-Carlo radiative transfer simulations were then performed for all these clusters, with line emissivities and oscillator strengths taken from the Astrophysical Plasma Emission Code (APEC) and ATOMDB\footnote{http://www.atomdb.org/index.php} database \citep{Foster2012}. The rms turbulent velocity is set to $\sigma_{\rm turb} = 100$ km/s. We compute the the PDFs from the central 30 kpc, corresponding to 0.5 -1 $r_c$ and spanning angles of 0.73 - 2.3 arcmin for these clusters. The spanning angles are comparable to the HPD $\sim 1.3$ arcmin of {\it Astro-H}. Fig \ref{fig:ccc_pdf} shows the comparisons between our model and simulations, where the dashed and dotted curves are when the electron and ion distributions are used as inputs. Our model is in good agreement with simulations when the $\beta$ model of the ion distribution is used as the input. The flux and variance ratios agree within 8\% and 4\%, respectively. As expected, the agreement becomes worse when the electron distribution is used as the input. 
In Fig. \ref{fig:ccc_pdf1}, we compare PDFs accumulated between 30-60 kpc. The discrepancy between the model and simulation is somewhat larger, but still perfectly acceptable. The prominent exception is A3526, where the ion and electron profiles differ significantly in the center. Here, the PDFs are significantly overestimated by our model in these outer regions if the $\beta$ model fit of the ion distribution is used as the input. These results can be understood as follows. RS affects the cluster center primarily as a source of effective absorption, by scattering photons out of the line of sight. All that matters in this case is the central line optical depth (as we shall see in \S\ref{sec:analytic}), which is set by the ion profile. However, further out, RS is also important as a source of re-emission, by scattering photons emitted from the cluster core back into the line of sight. While adopting the cuspy ion profile here correctly estimates $\tau(r)$, it incorrectly overestimates emission from the core $\propto n_{i} n_{e}$, thus predicting an overabundance of scattered photons. {We find that a simple scheme could be used to largely correct this bias. With the measured temperature and metallicity profile, we first compute the overestimated emission from the core. Then, assuming this emission is from a point source in the cluster center, we subtract their contributions from the line profile. The corrected line profiles, shown in Fig. \ref{fig:ccc_pdf} with thick dotted blue curves, follow the simulated profiles more closely. In particular, the agreement is now excellent for A3526. In order to apply this correction, a new calibration has been performed for a point source in the cluster center. For this calibration, we use our parametric form to fit line profiles enclosed in a radius $r$, instead of those from a differential area at $r$. The latter could not be well-fitted by our model, since the emission come from scattered photons. To apply the model, the original emission from the point source has to be included. The profile from an infinitesimal area should be obtained from the calibration by taking a derivative $dP/(2\pi r)dr$. This ``point-source" calibration is also publicly available on the internet.}


{Based on these results, we conclude that our fits are broadly applicable to both CCCs and NCCCs. Using the ion $\beta$ profiles generally provides more accurate results than using the electron $\beta$ profiles. Where they are observed to differ significantly (as in A3526), a ``point source" correction should be performed to correct line profiles originating from outside the core.}

\subsection{Error Estimates and Analysis of Full Line Complex}
\label{subsec:constraints}

In this section, we discuss statistical and systematic errors in the moments of the line profile introduced by RS. By applying our model to the full 6.7 keV Fe line complex, which harbors both optically thick and thin lines, we show that RS has a fairly mild effect on statistical uncertainty, but a large effect on systematic errors. Note that some fraction of the latter carries over if RS is imperfectly modeled. In the Gaussian model, the constraint on turbulence are solely driven by
the variance of the line. RS modifies the line profile, changing
both the height and its shape. So there are three sources of
constraints: the line dispersion, the total flux in the line, as well as deviations from Gaussianity. Although the last potentially contains extremely valuable information \citep{shang12}, the effects of RS can be more complicated there, and we defer a detailed investigation to future work. We therefore focus on the
first two constraints, which correspond to the zeroth and second moments of the line profile.  

\begin{figure}
\begin{tabular}{c}
\rotatebox{-90}{\resizebox{90mm}{!}{\includegraphics{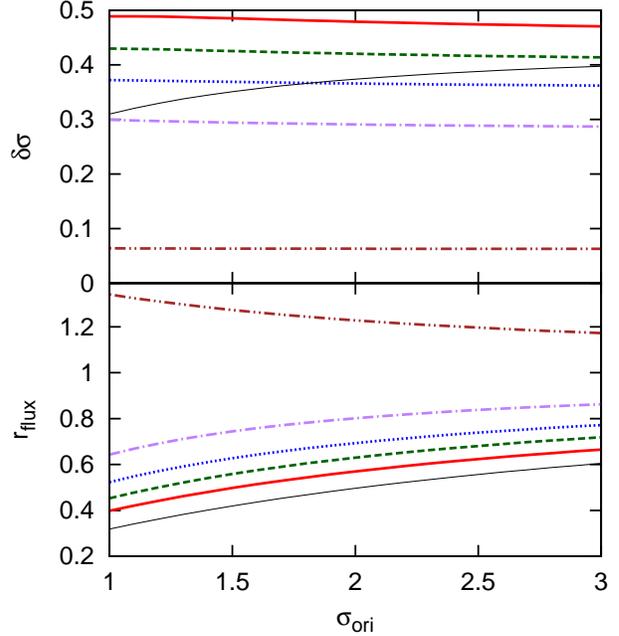}}}
\end{tabular}
\caption{Dispersion difference $\delta \sigma \equiv.
  \sigma_{RS}-\sigma_{\rm ori}$ and flux ratio as functions of 
  $\sigma_{\rm ori}$, the line width without RS. These lines all assume
  $\zeta=5$ and $\beta=0.6$. Photon are accumulated over rings with
  \{${\rm inner~radius, width}$\}=\{${0, 0.5r_c}$\} (solid red line), \{${0, r_c}$\}
  (dashed green line), \{$0,1.5r_c$\}(dotted blue line), \{$r_c,
  0.5r_c$\} (dot-dashed purple line) and \{$3r_c, 0.5r_c$\}
  (dot-dot-dashed brown line).} 
\label{fig:trend}
\end{figure}

It is easy to see why resonant scattering causes a mild increase in the statistical uncertainty of turbulent line broadening: it increases the dispersion of the line, and decreases the flux (or the number of photons) in a line. These effects are shows in Fig. \ref{fig:trend}, where different line types correspond to different radii and
width of photon collecting rings. Flux is mostly absorbed in the inner regions $r_{\rm flux} < 1$, and re-emitted in the outer regions $r_{\rm flux} > 1$. Note that the difference in dispersion between profiles with and without resonant scattering, $\delta \sigma = \sigma_{\rm RS} -\sigma_{\rm ori} \approx$ const, independent of $\sigma_{\rm ori}$. This allows us to write: 
\begin{equation}
\frac{\Delta \sigma_{\rm turb, RS}}{\Delta \sigma_{\rm turb, ori}} = \frac{1}{\sqrt{r_{\rm flux}}} \left( 1+ \frac{\delta \sigma}{\sigma_{\rm ori}} \right) 
\end{equation}
which is greater than unity, since $r_{\rm flux} < 1$ and $\delta \sigma > 0$. However, as we shall see, we expect errors to be dominated by systematic errors (due to imperfect modeling of resonant scattering, which has a large effect), rather than statistical errors (which are small, due to the large number of photons $n_{\rm ph} \sim 10^{4}$.

\begin{figure}
\begin{tabular}{c}
\rotatebox{-90}{\resizebox{100mm}{!}{\includegraphics{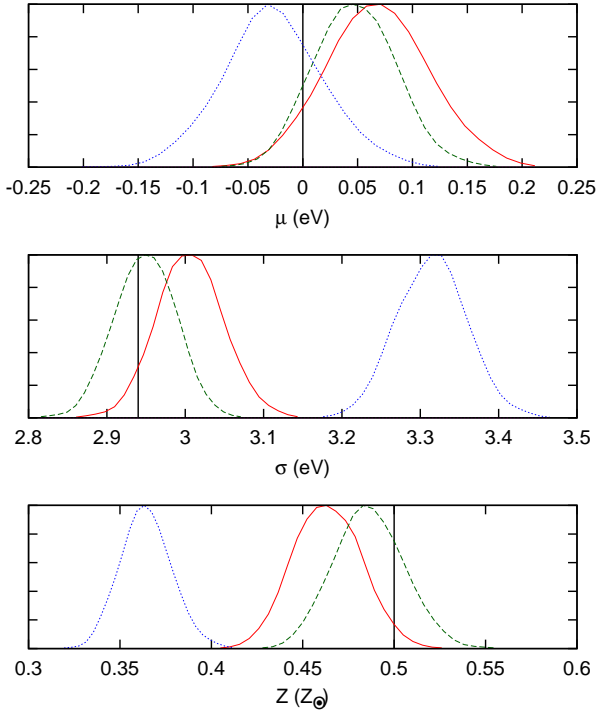}}}
\end{tabular}
\caption{1D marginalized posterior distributions for $\mu$, $\sigma$
  and $Z$. Solid red, dashed green and dotted blue curves correspond
  to case 0, 1 and 2, respectively. The vertical black lines indicates
the location of the input values. Ignoring resonant scattering results in significant biases in the recovered dispersions $\sigma$ and metallicities $Z$.}
\label{fig:fit}
\end{figure}

We now perform a fully realistic spectral line fit of the entire 6.7 keV Fe line complex. Besides allowing a fully accounting of the error budget, it highlights the systematic biases introduced by RS if it is ignored. It also serves as a final check of our model. 
We first
produce mock data of line profiles from the inner $30$ kpc of the
Perseus cluster, using the temperature, density and metallicity profiles described in \S~\ref{subsec:application}. The 1D RMS velocity is set
to $\sigma_{\rm turb} = 100$ km/s\footnote{This number is illustrative; 
  there are hints that turbulence in Perseus could be substantially stronger 
  \citep{Churazov2004}.}. We renormalize the profiles 
such that the photon counts in the Helium-like iron line is $10^4$,
roughly the expected counts of Astro-H from local bright clusters
\citep{Shang2012}. The counts in other lines and in the continuum are
assigned according to the APEC model
\citep{Foster2012}, assuming a metallicity of  
0.5 $Z_{\odot}$. Convolving with the telescope response (assuming FWHM = 5 eV), we produce mock spectra both with and without
RS. Most lines in the complex usually have very small optical depth,
except the Helium-like iron line. For simplicity, we here assume only
the Helium-like iron line is affected by the RS. We then use the
Gaussian and the new model to fit the spectra, adopting $\chi^2$
statistics (equation \ref{eqn:chisq}) and the Markov chain Monte Carlo algorithm implemented in
{\sc CosmoMC} \citep{Lewis2002}. There are four free parameters: the original photon counts in the Helium-like iron line $n_{ph}$,
the mean $\mu$, the original width $\sigma$ and the metallicity $Z$. We studied 3 cases:
the new model fitting the RS spectrum, the Gaussian model fitting the
Gaussian spectrum (unmodified by RS) and the Gaussian model fitting the RS spectrum. The
results are given in Table \ref{tbl:fit}; Fig. \ref{fig:fit} shows the marginalized posterior distributions. They show our model
correctly reproduces 
these quantities. The constraints are a little worse than those
in the case 1, as expected. In case 2 where the effects of RS are ignored, the parameters are significantly biased. The
line width is overestimated due to the additional broadening caused by
RS. The metallicity is on the other hand underestimated due to the
reduced counts in the line. An improved model like ours is therefore
imperative to correctly extract the underlying parameters.

\begin{table}
  \caption{Best-fit values and 68\% uncertainties from fitting 
    the mock spectra. Case 0 is when our model is applied to a RS
    spectrum, case 1 is when the Gaussian model is applied to a
    Gaussian spectrum (i.e., without RS), and case 2 is when the Gaussian model is applied
  to a RS spectrum. }
    \label{tbl:fit}
\begin{center}
\begin{tabular}{c c c c c}
\hline
\hline
& $n_{ph}~(\times 10^4)$ & $\mu$ (eV) & $\sigma$ (eV) & $Z~(Z_{\odot})$  \\
\hline
Input & 1 & 0 & 2.82 & 0.3 \\
\hline
Case 0 & $0.965_{-0.008}^{+0.008}$ & $0.069_{-0.045}^{+0.046}$ & $3.006_{-0.040}^{+0.041}$ & $0.464_{-0.018}^{+0.018}$ \\
Case 1 & $0.996_{-0.008}^{+0.008}$ & $0.048_{-0.038}^{+0.037}$ & $2.949_{-0.038}^{+0.039}$ & $0.486_{-0.019}^{+0.018}$\\
Case 2 & $0.836_{-0.008}^{+0.007}$ & $-0.033_{-0.043}^{+0.044}$ & $3.315_{-0.044}^{+0.043}$ & $0.364_{-0.014}^{+0.014}$\\
\hline
\end{tabular}
\end{center}
\end{table}

\subsection{Analytic Model for the Effects of RS} 
\label{sec:analytic}

In this section, we discuss an approximate analytic model for the modifications in the line profile due to resonant scattering. This drives both physical understanding, and allows very rapid estimates, particularly of important physical quantities such as modifications to the line height, variance, and flux. 

The previously calculated line profiles in the central regions have a characteristic shape: flat in the central regions, matching onto Doppler wings with relatively little change from the unscattered profile. This characteristic shape is a generic feature arising from radiative transfer of resonance line photons in moderately optically thick media, where the natural line wings are unimportant and only the Doppler core has significant optical depth. In this case, for line center optical depths $\tau_{0}$, for an initial line profile $J(x) = j e^{-x^{2}}/\sqrt{\pi}$ (i.e., $x \equiv \Delta E/\sqrt{2}\sigma$ in our notation), the final escaping line profile for isotropic scattering with no recoil is \citep{field59}:
\begin{equation}
\sqrt{\pi} \frac{J(x)}{j} = \frac{1-{\rm exp}(-\tau_{0}e^{-x^{2}})}{\tau_{0}} + 2 \int_{0}^{\tau_{0}e^{-x^{2}}} e^{-u} g_{2} \left( \frac{u}{l} \right) du
\label{eqn:RS_analytic_profile} 
\end{equation} 
where 
\begin{equation}
g_{n}(u) = \frac{1}{({\rm log} u^{-1})^{1/2}} \int_{0}^{({\rm log}u^{-1})^{1/2}} v^{n} e^{-v^{2}} dv. 
\end{equation}
We have found that this integral can be evaluated analytically: 
\begin{equation}
g_{2}(u) = \frac{\sqrt{\pi}}{4 y} {\rm erf}(y) - \frac{1}{2} e^{-y^{2}} 
\end{equation}
where $y=({\rm log} u^{-1})^{1/2}$. 

This profile was derived by considering the time development of the frequency distribution of a photon emitted in a nebula of infinite optical depth, but \citet{field59} showed that it can also be applied with reasonable approximation to a nebula with finite optical depth $\tau_{0}$. Focusing on the first term, we see that there is a critical frequency:
\begin{equation}
\tau_{\nu} = 1 \Rightarrow x_{c} = ({\rm log} \, \tau_{0})^{1/2} 
\end{equation}
at which the wings of the profile become optically thin, and photons can escape freely. Indeed, for $x > x_{\rm c}$, the derivative of the profile is approximately $-2x e^{-x^{2}}/\sqrt{\pi}$, its initial value, while for $ x < x_{\rm c}$, the derivative is ${\rm exp}(-\tau_{0} e^{-x^{2}})$, and hence quickly approaches zero. Thus, the profile is flat for $x < x_{\rm c}$, and relatively unchanged from the initial profile for $x > x_{\rm c}$ --- exactly what we have seen in our Monte-Carlo simulations. The escape probability per scattering is erfc($\sqrt{{\rm ln} \, \tau_{0}}$), and the mean number of scatterings is $\langle N \rangle = 1/{\rm erfc}(\sqrt{{\rm ln} \, \tau_{0}})$. 

Equation \ref{eqn:RS_analytic_profile} was obtained by explicit solution of the integro-differential equation of radiative transfer. Nonetheless, we can intuitively understand this line shape from the form of the redistribution function $r(x,x^{\prime})$, which gives the probability that radiation of frequency $x$ will be scattered into frequency $x^{\prime}$. For isotropic scattering, it is \citep{zanstra49,field59}:
\begin{equation}
r(x,x^{\prime}) = \int_{|\tilde{x}|}^{\infty} e^{-v^{2}}dv; \ |\tilde{x}|={\rm max}(|x|,|x^{\prime}|).
\end{equation} 
where $v$ is in units of the 1D velocity dispersion $\sigma_{v}$; note that $r(x,x^{\prime})$ is symmetric. Physically, we can understand this expression from the fact that the speed of an atom must be larger than both $x$ and $x^{\prime}$ to scatter $x$ into $x^{\prime}$; as long as this is true, then for isotropic scattering the scattering probability is independent of $x$ and $x^{\prime}$. Thus, for a photon of frequency $x$, the redistribution function is {\it constant} (independent of $x^{\prime})$, until $x^{\prime}=x$, at which point $r(x,x^{\prime})$ plummets rapidly. We can apply this to understand the form of the line profile. In the optically thick core $x < x_{c}$, photons are trapped and scatter repeatedly. Due to the form of the redistribution function, the scattered photons have a flat profile between $-x$ and $x$, with a finite chance to escape to $x^{\prime} > x$. One can think of the photons as being trapped between reflecting boundaries at $-x$ and $x$, which have some finite probability of transmission. Thus, the Doppler profile in the core is erased to form a flat profile. Photons generally escape only once they hit a rare fast moving atom, and are re-emitted with $x > x_{c}$. Photons initially emitted in the optically thin wings $x>x_{c}$ of course travel unimpeded towards us, and the line wings are unaltered from the initial profile. 

Besides redistribution in frequency, RS also causes spatial diffusion: photons migrate from the optically thick cluster core to the optically thin cluster outskirts, causing the surface brightness profile to be more heavily weighted toward the optically thin 'photosphere' than the emission profile (in the absence of any destruction processes, the total line flux from the cluster is, of course, conserved; the total area under the curve in equation (\ref{eqn:RS_analytic_profile}) is unity). This can be clearly seen in Fig \ref{fig:trend}, where the flux ratio in the line with and without resonance scattering is $r_{\rm flux} < 1$ in the inner regions and $r_{\rm flux} > 1$ in the outer regions. Also, from Table \ref{tbl:result}, the parameter $g_{1}$, which parametrizes the strength of re-emission, increases with radius. We therefore have to be careful when comparing the analytic formula in equation (\ref{eqn:RS_analytic_profile}) to simulation results. Let us focus on the fate of photons which are generated in the inner cluster core; the fraction of photons which originate in the cluster outskirts is small. When considering the emission from radial line profiles $P(r)$, in the inner regions the photons which are scattered out of the line of sight (to be re-emitted in the outer regions) are effectively `absorbed', since they do not contribute to emission from that radius. Thus, {\it only the first term in equation (\ref{eqn:RS_analytic_profile}) should be used}. However, when considering the observed line profile emerging from the cluster as a whole (which involves an integral over all radii), both terms in equation (\ref{eqn:RS_analytic_profile}) should be used, since in this case the photon flux is conserved. 


This analytic formula requires the optical depth at line center. Note that this is a spatially varying quantity, since a photon from the near side of a cluster scatters less than a photon from the far side of the cluster. For a shell at radius $r$, the correct optical depth to use is: 
\begin{eqnarray}
&&\tau_{\rm eff}(r) = -{\rm log}\left(\frac{1}{4\pi}\int d{\bf \hat{n}} \int_{0}^{\infty} dl \ {\rm exp}[-n({\bf n},r) \sigma l]\right)\\\nonumber
&=& -{\rm log}\left(\frac{1}{2}\int d cos(\theta) \int_{0}^{\infty} dl \ {\rm exp}[-n(\sqrt{l^2+r^2-2r l cos(\theta)}) \sigma l]\right) 
\end{eqnarray} 
For shells close to the cluster center, the correction is small, and a simple integral from $r$ to infinity suffices.  


Fig. \ref{fig:example} compares the analytic profile to simulation results; they show good agreement. An analytic profile allows simple statistics such as modifications to line height, variance and flux due to resonant scattering to be quickly estimated; we find that these reproduce the trends seen in Fig \ref{fig:trend}, and are accurate to $\sim 20-30\%$. 

We stress that the analytic profile is only approximate; quantitative conclusions should always been drawn from radiative transfer simulations. The analytic profile is accurate near the cluster center and for the spectrum emerging from the cluster as a whole, but modeling the radial trends seen in our simulations (which reflect photons from the cluster core scattered into the line of sight) is beyond the scope of this paper. The analytic approach also makes certain restrictive assumptions, though these turn out not to matter significantly. Equation (\ref{eqn:RS_analytic_profile}) is derived under the assumption that $\tau_{0} \gg 1$, though in practice it should be reasonably accurate for moderate $\tau_{0} > 1$. Note also that equation (\ref{eqn:RS_analytic_profile}) also assumes isotropic scattering, whereas for instance our fiducial 6.7 keV He-like Fe is pure dipole scattering. Dipole scattering has a different redistribution function $r(x,x^{\prime})$ from isotropic scattering; there is a correlation between the ingoing and outgoing photon frequencies, since the scattering tends to be in the same direction as absorption. In practice, we find from our Monte-Carlo simulations that calculations which assume isotropic and dipole scattering do not differ significantly. Overall, the analytic approach is useful for physical insight and quick `rule of thumb' estimates.  

\section{Summary}
\label{sec:summary}

Upcoming telescopes will enable us, for the first time, to measure ICM turbulence through Doppler broadening of
emission lines. However, the brightest lines with the highest signal to noise also tend to be mildly optically thick. Resonance scattering of emission lines needs to be taken into account, as it
alters the line flux and more importantly, increases the line widths used to measure turbulence. Naive measurements which ignore this would give biased results. To date, characterizing and eliminating this bias has not been quantitatively addressed. 
In this paper, we proposed a new model for the emission line
profile, accounting for the effects of resonant scattering. This model assumes that the ions and electrons both obey $\beta$ profiles, and works best in the limit where these profiles are fairly similar in shape in the core. 

Motivated by cosmological simulations of galaxy clusters, we assume that the velocity field in galaxy clusters is adequately characterized by a power spectrum alone, and show that as long as the outer scale is smaller than the core size, for a given rms velocity, our results are not sensitive to the assumed power spectrum. We run Monte Carlo radiative transfer simulations and show that our model provides an excellent fit with only 3 parameters, allowing the effects of resonant scattering and turbulent broadening to be disentangled. We provide fitting formulae and tables for model parameters which will allow rapid fits to observed line profiles. As a test, we run MCMC simulations of the entire Fe line complex for the Perseus cluster, and show our model provides unbiased estimates of turbulent broadening with relatively small error bars, where neglecting resonant scattering clearly results in biased estimates of turbulent widths and metallicities. Finally, we show that the physical origin of the calculated line profiles ---which are typically flat in a central core and have relatively unmodified wings -- can be understood analytically. Our model assumes that the electron profile (which affects the emissivity profile) and the ion profile (which affects both the emissivity and optical depth profile) have relatively similar $\beta$ profile shapes in the core; where they differ, the ion profile should be used. When they differ significantly, a ``point source" correction should be used to correct emission originating from outside the core. In this way, CCCs are well fitted by our model as well as NCCCs.

Several extensions and improvements to this work are possible. Our discussions have focussed on the He-like 6.7 keV Fe line; detailed considerations of other lines would be worthwhile. Secondly, as we discussed in \citet{Shang2012},
distinctive large scale modes of gas motions arise in clusters. These
modes could potentially be separated and individually constrained
by mixture modeling. Such mode separation could
be very useful separating bulk motions (e.g., gas sloshing in clusters with cold fronts) from turbulence, and discerning volume filling
fraction of turbulence. The effect of resonant scattering has yet to be taken into account in mixture models, which exploit non-Gaussian features in the line profile. 

\vspace{-0.5\baselineskip}

\section*{Acknowledgments}

We acknowledge NASA grant NNX12AG73G for support. SPO thanks UCLA for 
hospitality. We thank Irina Zhuravleva for remarks in a referee report which stimulated this work. 


\bibliography{library,rs,master_references}
\appendix
\section[]{Best-fit model parameters}
\begin{table}
  \caption{Best-fit model parameters for different $\beta$ and $r$.}
    \label{tbl:result}
\begin{center}
\begin{tabular}{c c c c}
\hline
\hline
$r/r_c$ & $p_1$ & $p_2$ & $h$\\
\hline
\multicolumn{4}{|c|}{$\beta$ = 0.5}\\
\hline
0.05 & $0.098 \pm 0.027$ & $0.615 \pm 0.028$ & $1.460 \pm 0.224$\\
0.55 & $0.092 \pm 0.008$ & $0.489 \pm 0.009$ & $1.359 \pm 0.074$\\
1.05 & $0.122 \pm 0.007$ & $0.396 \pm 0.007$ & $1.089 \pm 0.051$\\
1.55 & $0.124 \pm 0.008$ & $0.301 \pm 0.007$ & $1.100 \pm 0.049$\\
2.05 & $0.141 \pm 0.008$ & $0.237 \pm 0.008$ & $1.012 \pm 0.047$\\
2.55 & $0.153 \pm 0.009$ & $0.185 \pm 0.008$ & $0.955 \pm 0.047$\\
3.05 & $0.154 \pm 0.009$ & $0.140 \pm 0.009$ & $0.942 \pm 0.049$\\
3.55 & $0.153 \pm 0.010$ & $0.105 \pm 0.009$ & $0.921 \pm 0.053$\\
4.05 & $0.165 \pm 0.010$ & $0.082 \pm 0.009$ & $0.862 \pm 0.054$\\
4.55 & $0.169 \pm 0.011$ & $0.067 \pm 0.010$ & $0.844 \pm 0.056$\\
5.05 & $0.164 \pm 0.012$ & $0.048 \pm 0.011$ & $0.933 \pm 0.060$\\
5.55 & $0.167 \pm 0.012$ & $0.035 \pm 0.011$ & $0.912 \pm 0.061$\\
6.05 & $0.140 \pm 0.013$ & $-0.005 \pm 0.012$ & $0.907 \pm 0.077$\\
6.55 & $0.153 \pm 0.014$ & $0.002 \pm 0.013$ & $0.922 \pm 0.076$\\
7.05 & $0.164 \pm 0.015$ & $0.006 \pm 0.013$ & $0.938 \pm 0.074$\\
\hline
\multicolumn{4}{|c|}{$\beta$ = 0.6}\\
\hline
0.05 & $0.074 \pm 0.019$ & $0.550 \pm 0.021$ & $1.541 \pm 0.213$\\
0.55 & $0.095 \pm 0.007$ & $0.464 \pm 0.007$ & $1.274 \pm 0.056$\\
1.05 & $0.128 \pm 0.006$ & $0.352 \pm 0.006$ & $1.103 \pm 0.040$\\
1.55 & $0.147 \pm 0.007$ & $0.256 \pm 0.007$ & $1.002 \pm 0.039$\\
2.05 & $0.161 \pm 0.008$ & $0.178 \pm 0.007$ & $0.946 \pm 0.041$\\
2.55 & $0.172 \pm 0.009$ & $0.123 \pm 0.008$ & $0.861 \pm 0.044$\\
3.05 & $0.175 \pm 0.010$ & $0.082 \pm 0.009$ & $0.878 \pm 0.047$\\
3.55 & $0.174 \pm 0.011$ & $0.051 \pm 0.010$ & $0.850 \pm 0.053$\\
4.05 & $0.165 \pm 0.012$ & $0.012 \pm 0.011$ & $0.856 \pm 0.062$\\
4.55 & $0.169 \pm 0.013$ & $-0.006 \pm 0.012$ & $0.863 \pm 0.065$\\
5.05 & $0.165 \pm 0.014$ & $-0.030 \pm 0.013$ & $0.874 \pm 0.071$\\
5.55 & $0.154 \pm 0.015$ & $-0.047 \pm 0.014$ & $0.874 \pm 0.082$\\
6.05 & $0.173 \pm 0.016$ & $-0.044 \pm 0.015$ & $0.820 \pm 0.081$\\
6.55 & $0.174 \pm 0.018$ & $-0.040 \pm 0.016$ & $0.931 \pm 0.085$\\
7.05 & $0.182 \pm 0.019$ & $-0.040 \pm 0.017$ & $0.860 \pm 0.090$\\
\hline
\multicolumn{4}{|c|}{$\beta$ = 0.7}\\
\hline
0.05 & $0.088 \pm 0.015$ & $0.545 \pm 0.016$ & $1.411 \pm 0.145$\\
0.55 & $0.100 \pm 0.005$ & $0.439 \pm 0.006$ & $1.256 \pm 0.045$\\
1.05 & $0.140 \pm 0.006$ & $0.308 \pm 0.005$ & $1.028 \pm 0.033$\\
1.55 & $0.166 \pm 0.006$ & $0.197 \pm 0.006$ & $0.904 \pm 0.034$\\
2.05 & $0.175 \pm 0.008$ & $0.114 \pm 0.007$ & $0.886 \pm 0.038$\\
2.55 & $0.179 \pm 0.009$ & $0.054 \pm 0.008$ & $0.906 \pm 0.044$\\
3.05 & $0.195 \pm 0.011$ & $0.025 \pm 0.009$ & $0.855 \pm 0.048$\\
3.55 & $0.203 \pm 0.012$ & $-0.003 \pm 0.011$ & $0.826 \pm 0.053$\\
4.05 & $0.210 \pm 0.014$ & $-0.020 \pm 0.012$ & $0.852 \pm 0.059$\\
4.55 & $0.210 \pm 0.016$ & $-0.048 \pm 0.014$ & $0.791 \pm 0.069$\\
5.05 & $0.206 \pm 0.018$ & $-0.060 \pm 0.015$ & $0.833 \pm 0.075$\\
5.55 & $0.204 \pm 0.020$ & $-0.082 \pm 0.017$ & $0.851 \pm 0.085$\\
6.05 & $0.208 \pm 0.022$ & $-0.083 \pm 0.019$ & $0.783 \pm 0.093$\\
6.55 & $0.205 \pm 0.025$ & $-0.091 \pm 0.022$ & $0.880 \pm 0.103$\\
7.05 & $0.240 \pm 0.028$ & $-0.085 \pm 0.024$ & $0.742 \pm 0.106$\\
\hline
\end{tabular}
\end{center}
\end{table}
\begin{table}
  \caption{Continued from Table \ref{tbl:result}.}
\begin{center}
\label{tbl:result1}
\begin{tabular}{c c c c}
\hline
\hline
$r/r_c$ & $p_1$ & $p_2$ & $h$\\
\hline
\multicolumn{4}{|c|}{$\beta$ = 0.8}\\
\hline
0.05 & $0.105 \pm 0.013$ & $0.540 \pm 0.013$ & $1.277 \pm 0.101$\\
0.55 & $0.113 \pm 0.005$ & $0.416 \pm 0.005$ & $1.218 \pm 0.035$\\
1.05 & $0.156 \pm 0.005$ & $0.267 \pm 0.005$ & $0.995 \pm 0.028$\\
1.55 & $0.173 \pm 0.006$ & $0.139 \pm 0.006$ & $0.889 \pm 0.032$\\
2.05 & $0.196 \pm 0.008$ & $0.058 \pm 0.007$ & $0.808 \pm 0.037$\\
2.55 & $0.203 \pm 0.010$ & $-0.005 \pm 0.009$ & $0.816 \pm 0.046$\\
3.05 & $0.236 \pm 0.013$ & $-0.022 \pm 0.011$ & $0.807 \pm 0.049$\\
3.55 & $0.240 \pm 0.015$ & $-0.062 \pm 0.013$ & $0.772 \pm 0.059$\\
4.05 & $0.235 \pm 0.018$ & $-0.097 \pm 0.015$ & $0.737 \pm 0.070$\\
4.55 & $0.271 \pm 0.021$ & $-0.086 \pm 0.017$ & $0.762 \pm 0.071$\\
5.05 & $0.258 \pm 0.025$ & $-0.122 \pm 0.020$ & $0.821 \pm 0.085$\\
5.55 & $0.340 \pm 0.030$ & $-0.065 \pm 0.022$ & $0.827 \pm 0.078$\\
6.05 & $0.309 \pm 0.035$ & $-0.111 \pm 0.027$ & $0.818 \pm 0.099$\\
6.55 & $0.323 \pm 0.040$ & $-0.115 \pm 0.031$ & $0.856 \pm 0.104$\\
7.05 & $0.391 \pm 0.048$ & $-0.098 \pm 0.034$ & $0.767 \pm 0.109$\\
\hline
\multicolumn{4}{|c|}{$\beta$ = 0.9}\\
\hline
0.05 & $0.099 \pm 0.011$ & $0.521 \pm 0.012$ & $1.325 \pm 0.094$\\
0.55 & $0.120 \pm 0.004$ & $0.389 \pm 0.004$ & $1.164 \pm 0.030$\\
1.05 & $0.162 \pm 0.005$ & $0.221 \pm 0.005$ & $0.972 \pm 0.026$\\
1.55 & $0.194 \pm 0.007$ & $0.099 \pm 0.006$ & $0.862 \pm 0.030$\\
2.05 & $0.217 \pm 0.009$ & $0.011 \pm 0.008$ & $0.812 \pm 0.038$\\
2.55 & $0.254 \pm 0.012$ & $-0.040 \pm 0.010$ & $0.715 \pm 0.045$\\
3.05 & $0.286 \pm 0.016$ & $-0.058 \pm 0.013$ & $0.821 \pm 0.050$\\
3.55 & $0.324 \pm 0.020$ & $-0.079 \pm 0.016$ & $0.757 \pm 0.058$\\
4.05 & $0.330 \pm 0.025$ & $-0.121 \pm 0.019$ & $0.713 \pm 0.072$\\
4.55 & $0.346 \pm 0.032$ & $-0.141 \pm 0.024$ & $0.795 \pm 0.084$\\
5.05 & $0.367 \pm 0.039$ & $-0.137 \pm 0.028$ & $0.886 \pm 0.095$\\
5.55 & $0.475 \pm 0.048$ & $-0.149 \pm 0.032$ & $0.629 \pm 0.096$\\
6.05 & $0.492 \pm 0.058$ & $-0.181 \pm 0.037$ & $0.531 \pm 0.119$\\
6.55 & $0.643 \pm 0.075$ & $-0.119 \pm 0.043$ & $0.817 \pm 0.105$\\
7.05 & $0.749 \pm 0.095$ & $-0.124 \pm 0.050$ & $0.592 \pm 0.125$\\
\hline
\multicolumn{4}{|c|}{$\beta$ = 1.0}\\
\hline
0.05 & $0.109 \pm 0.010$ & $0.518 \pm 0.010$ & $1.302 \pm 0.076$\\
0.55 & $0.122 \pm 0.004$ & $0.361 \pm 0.004$ & $1.147 \pm 0.027$\\
1.05 & $0.176 \pm 0.005$ & $0.184 \pm 0.004$ & $0.927 \pm 0.024$\\
1.55 & $0.209 \pm 0.007$ & $0.044 \pm 0.006$ & $0.831 \pm 0.030$\\
2.05 & $0.255 \pm 0.010$ & $-0.030 \pm 0.009$ & $0.773 \pm 0.038$\\
2.55 & $0.300 \pm 0.015$ & $-0.080 \pm 0.012$ & $0.764 \pm 0.046$\\
3.05 & $0.333 \pm 0.020$ & $-0.119 \pm 0.015$ & $0.764 \pm 0.058$\\
3.55 & $0.390 \pm 0.027$ & $-0.140 \pm 0.020$ & $0.691 \pm 0.068$\\
4.05 & $0.536 \pm 0.039$ & $-0.108 \pm 0.025$ & $0.709 \pm 0.071$\\
4.55 & $0.519 \pm 0.048$ & $-0.164 \pm 0.030$ & $0.667 \pm 0.089$\\
5.05 & $0.748 \pm 0.067$ & $-0.137 \pm 0.036$ & $0.563 \pm 0.090$\\
5.55 & $0.768 \pm 0.089$ & $-0.156 \pm 0.045$ & $0.688 \pm 0.119$\\
6.05 & $0.868 \pm 0.123$ & $-0.229 \pm 0.058$ & $0.559 \pm 0.150$\\
6.55 & $1.222 \pm 0.169$ & $-0.137 \pm 0.060$ & $0.482 \pm 0.160$\\
7.05 & $1.671 \pm 0.266$ & $-0.079 \pm 0.067$ & $0.517 \pm 0.185$\\
\hline
\end{tabular}
\end{center}
\end{table}


\label{lastpage}
\end{document}